\documentclass[twoside,a4paper,11pt]{sea22}
\usepackage{graphicx}
\usepackage{hyperref}

\begin{document}
\pagenumbering{arabic}
\pagestyle{myheadings}
\thispagestyle{empty}
{\flushleft\includegraphics[width=\textwidth,bb=58 650 590 680]{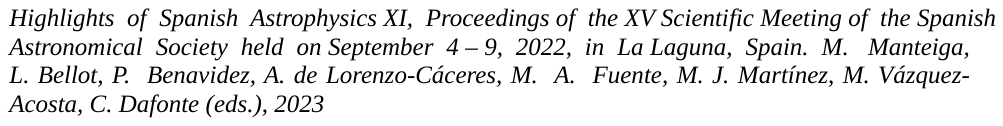}}
\vspace*{0.2cm}
\begin{flushleft}
{\bf {\LARGE
%
Astro+: Design, construction, and scientific exploitation of a large-scale massive star spectroscopic database
%
}\\
\vspace*{1cm}
%
Klaus R\"ubke$^{1}$,
 Amparo Marco$^{1}$,
 Ignacio Negueruela$^{2}$,
 Artemio Herrero$^{3,4}$,
 Sergio Simon-Diaz$^{3,4}$,
 Hugo Tabernero$^{5}$
 and  Lee Patrick$^{2,5,6}$
%
}\\
\vspace*{0.5cm}
%
{\footnotesize

$^{1}$
Departamento de Física, Ingeniería de Sistemas y Teoría de la Señal, Universidad de Alicante, Carretera de San Vicente s/n, E03690, San Vicente del Raspeig, Spain e-mail : klaus.rubke@ua.es
\\
$^{2}$
Departamento de Física Aplicada, Facultad de Ciencias, Universidad de Alicante, Carretera de San Vicente s/n, E03690, San Vicente del Raspeig, Spain
\\
${^3}$
Instituto de Astrofísica de Canarias, Vía Láctea s/n, E38200, La Laguna, Tenerife, Spain
\\
${^4}$
Universidad de La Laguna, Departamento de Astrofísica, E38206 La Laguna, Tenerife, Spain
\\
${^5}$
Centro de Astrobiología, CSIC-INTA. Campus ESAC. C. bajo del castillo s/n. E-28 692 Villanueva de la Cañada, Madrid, Spain
\\
${^6}$
School of Physical Sciences, The Open University, Walton Hall, Milton Keynes MK7 6AA, UK
}
%
\end{flushleft}
%
\markboth{
Astro+
}{ 
%
R\"ubke, K. et al. 
%
}
\thispagestyle{empty}
\vspace*{0.4cm}
\begin{minipage}[l]{0.09\textwidth}
\ 
\end{minipage}
\begin{minipage}[r]{0.9\textwidth}
\vspace{1cm}
\section*{Abstract}{\small
%
Massive stars condition the evolution of the interstellar medium by the amount of energy released during their lives and especially by their deaths as supernova explosions. The vast amounts of spectroscopic data for massive stars provided by previous and existing instruments on ground-based and space-based telescopes have already saturated our capability to process them by the use of human routines. As a consequence, there is a pressing need for machine-assisted tools to help handle incoming data. To this end, we present the development of a massive star spectroscopic multiwavelength interactive database designed for scientific research and a fully automatic stellar parameter determination tool. Here we show the preliminary results of the application of these tools to optical spectra of O-type stars.
%
\normalsize}
\end{minipage}
%
%
\section{Introduction}
Massive stars play a very important role in the evolution and formation of galaxies. They possess strong winds energising their neighbourhood and through their deaths as a supernova explosion, enrich the interstellar medium (ISM) with heavier materials. Having a centralized database of spectra of massive stars is the goal of this research, as it is crucial to have a place where homogenous procedures could study their properties.

In this matter, we present the first multi-wavelength database dedicated to massive star spectra, covering from Xrays to far infrared, \href{https://astroplus.ua.es}{Astro+}, both for young OB-type and evolved Red Supergiants (RSGs), of K and M spectral types. The purpose of this project is to host as many spectra of massive stars as possible. To this end, we provide to the scientific community, a tool to upload spectra through a user-friendly interface, so those who would like to contribute to our project can upload spectra to our database via the website {\bf https://astroplus.ua.es/}. We allow either FITS or ASCII formats.

The idea of having a database of massive star spectra, covering everything from young to evolved stars, provides us with a place to study and deeply understand the evolution of these stars and hence of the galaxy that hosts them. The first step in this process is to make a correct identification and characterization of the spectrum before further analysis can lead to the extraction of stellar parameters. To this aim, we have concentrated on creating an automation tool, \textit{HiLineThere}, for the correct identification of the type of star, differentiating whether the spectrum meets the properties corresponding to an OB or RSG star, and then determining stellar parameters depending on the type of star.

\section{The Database}

We have created a web application which is divided into three main components, the insertion, the search, and finally the analysis, which is running in the background, and it will be explained in more detail in section \ref{Science}.

The insertion of spectra is achieved via an interactive interface that greatly facilitates the correct upload of the data by any user. This versatile interface allows for uploading either one or many stars simultaneously, depending on their format. The platform accepts a multi-variety of FITS file formats and ASCII data.

More in detail, the upload of a FITS file is divided into four steps:

\begin{itemize}
    \item The user must provide the minimum mandatory information on the file. This includes the unit in which wavelength is expressed, if the spectrum is either not-normalized, normalized or flux calibrated, and the headers that will be extracted from the FITS file, such as Object Name, Right Ascension (RA), Declination (DEC), Observation Date, Telescope, Instrument and Exposure Time.
    \item The user must specify how the wavelength should be extracted; for example, if the wavelength is an array embedded in the data or if it must be calculated using CRVAL and CDELT1.
    \item The user must specify how the flux and the error of the flux, if they exist, should be extracted; for example, if this is a binary data or an array 
    \item Selecting the files to be uploaded to this corresponding configuration.
\end{itemize}


The upload of an ASCII file is straightforward but requires more preparation. This is due to the fact that a CSV file must be filled. In the file, the user must specify the name of the file where the spectrum is and the star header's mandatory parameters. The first step will ask for the unit of the wavelength if the spectrum is normalized, and the second step is where the header CSV file and the spectrum/s can be uploaded.

During the upload, we have provided the user with the possibility to check and correct any field or configuration, if needed. It is in this part where the uploaded files are standardised in Astro+'s own standard FITS format and are correlated with the GAIA and SIMBAD databases through their respective APIs. The last step to be completed by the user is to send the spectrum with all the parameters checked to the administrator for final review and subsequently transfer it to the final database.

The role of the Astro+ administrator is to check that the spectrum is correctly loaded and configured, e.g. that the wavelength is correct, that the resolution is that of the instrument, or that the different name was correctly acquired from SIMBAD or GAIA, when is actually not in those databases we have provided the option to classify the spectrum as new (not yet catalogued). If any of these items is not correct, the administrator will send back the spectrum for corrections.

After approval, the star will be available in the search engine of our application, where it can be searched either by name, coordinates or a more advanced search by SQL command, where further star parameters will be available.

\section{Science/Artificial Inteligence }\label{Science}
The possibility and capability to gather stellar spectra in a single repository give us an excellent position to develop tools for analysing the data. Although the idea of having full automation of both spectrum identification/characterisation and stellar parameter extraction sounds quite ambitious, we have created \textit{HiLineThere}, a fully automatic tool for characterisation and parameter determination. This tool considers standard diagnostic methods based on the literature and the working group members' and collaborators' expertise.

\begin{table}[!ht]
\label{table:1}

\caption{Diagnostic lines used by Astro+. Lines used for the determination of RV are marked with X, while the lines used to determine $v\sin\,i$ are numbered according to priority.}
\begin{center}
\footnotesize
\begin{tabular}[t]{llll}\hline
Line      & Lambda [\AA]  & Vrad  &Vsini \\\hline\hline
HALPHA    & 6562.80 &   & \\
HBETA     & 4861.33 &   & \\
HGAMMA    & 4340.46 &   & \\
HDELTA    & 4101.74 &   & \\
HEPSILON  & 3970.07 &   & \\
HEI4026   & 4026.19 & X & \\
HEI4387   & 4387.93 & X & \\
HEI4471   & 4471.47 & X & \\
HEI4713   & 4713.16 &   & \\
HEI4922   & 4921.93 &   & \\
HEI5876   & 5875.62 & X & \\
HEI6678   & 6678.15 &   & \\
HEII4200  & 4199.83 &   & \\
\vdots & \vdots & \vdots & \vdots  \\\hline
\end{tabular} 
\begin{tabular}[t]{llll}\hline
Line      & Lambda  [\AA]& Vrad  &Vsini \\\hline\hline
\vdots & \vdots & \vdots  & \vdots \\
HEII4541  & 4541.59 & X & 4\\
HEII4686  & 4685.71 &   & \\
HEII5411  & 5411.52 & X & 3 \\
HEII6527  & 6527.00    &   & \\
OIII5592  & 5592.37 & X & 1\\
SIII4128  & 4128.07 &   & \\
SIII4131  & 4130.89 &   & \\
SIII6347  & 6347.11 &   & \\
SIII6371  & 6371.37 &   & \\
SIIII4552 & 4552.00    & X & 2\\
SIIV2089  & 4088.85 &   & \\
SIIV4116  & 4116.10 &   & \\
MGII4481  & 4481.00    & X & 5\\
CII4267   & 4267.00    & X & 6\\ \hline
\end{tabular}
\end{center}
\end{table}

\textit{HiLineThere} is mainly divided into two main routines, characterisation and classification routine, from where we can check the presence or absence of diagnostic lines to identify the type of the star, and the Best Model fitting routine, where the spectrum is compared to a grid of synthetic models by using the $\chi^2$ technique, a simple but effective method to compare line profiles, in which the minimum of the sum of all lines, corresponds to the best fitting model, and the parameters used to generate this stellar spectrum are associated to the observed spectrum.

\subsection{Characterisation and classification}
This routine is based on iterations, which stop when the parameter $V\sin\,i$ converges, as explained later in this section. It starts by using a series of characteristic lines for OB-type stars (see Table \ref{table:1}). Using the laboratory centre of each line, we establish a search range of $\pm 300$ [Km/s] in the spectrum, as the radial velocity (RV) is not always corrected, and try to detect the presence of the line by attempting a line profile fit. We are considering in each iteration the standard procedures of renormalising and cleaning, before trying to find inflection points and finally trying to fit a gaussian and a pseudo-voigt profile within this range. 

If a line profile is identified, the following checks are performed:
\begin{itemize}
    \item If it has an equivalent width (EW) greater than 0.4 \AA, it is considered as detected. 
    \item It is considered to be absorption or emission based on the sign of EW.
    \item It is retained as detected if the centre displacement is within the $100$ [km/s] limit of the calculated RV, this is explained in the next paragraph (this is not considered in the first iteration).
    \item It is considered as quality detected if, in addition, the maximum or minimum of the line is greater than 3$\sigma$ of the flux selected in the range of the line.
\end{itemize}

As we mentioned, this process iterates three or more times. The first two iterations try to detect lines used to determine the RV of the spectrum using the centre of the detected lines and averaging them to obtain a general value used on the next iteration as input. In the third or higher iteration, the program uses one of the rotational velocity diagnostic lines to determine the projected rotational velocity  $V\sin\, i$. 

The method of determining $V\sin\, i$ follows the theory of \cite{Carroll1928}, \cite{Carroll1933a}, \cite{Carroll1933b}, \cite{Dravins1990}, \cite{Ramirez-Agudelo2013}, \cite{Reiners2003}, \cite{Reiners2001}, \cite{ssimon2007} and \cite{ssimon2006}, by using the fourier-transformation of the profile of the line to create the power spectrum from which the first zeros of the derivate corresponds to the pure rotational velocity of the star.

As the resolution limits the accuracy on the $V\sin\, i$ determination, we have set the program to stop when the velocity converges to less than a third of the speed as a function of the resolution ($\Delta V\sin\,i < \left(\frac{c_{light}}{Resolution}\right )/3$), and the determination uses the same line as the previous iteration, as the line used to determine $v\sin \, i$ may not be the same in the iteration.

After confirming the existence and good quality of the diagnostic lines, another subroutine \textit{isblue} determines if the star is an OB star, an RSG candidate star or an intermediate type star, based on counting the lines found. This counting is done separately for the H, HeI and HeII lines. For example, if a star has no detected HeI or HeII lines, it  will be considered as an RSG candidate.

\subsection{Best model fitting}

If the star is labelled as OB (blue path) or RSG (red path, not discussed in this article), it initiates the last part of the automation. The blue path uses the $\chi^2$ technique (\cite{ssimon2011}, \cite{ssimon2007},  \cite{ssimon2014} and \cite{ssimon2015}) to find the best model based on a synthetic grid of spectral model generated by the FASTWIND code (\cite{puls2005} and \cite{santolaya97}), obtaining the most critical stellar parameter, such as the effective temperature ($T_{\mathrm{eff}}$) and the gravity, expressed in log scale ($\log \, g$), in a completely automatic way for this type of star.

\section{Preliminary results}

\begin{figure}[!ht]
\centering 
\includegraphics[width=0.32\textwidth]{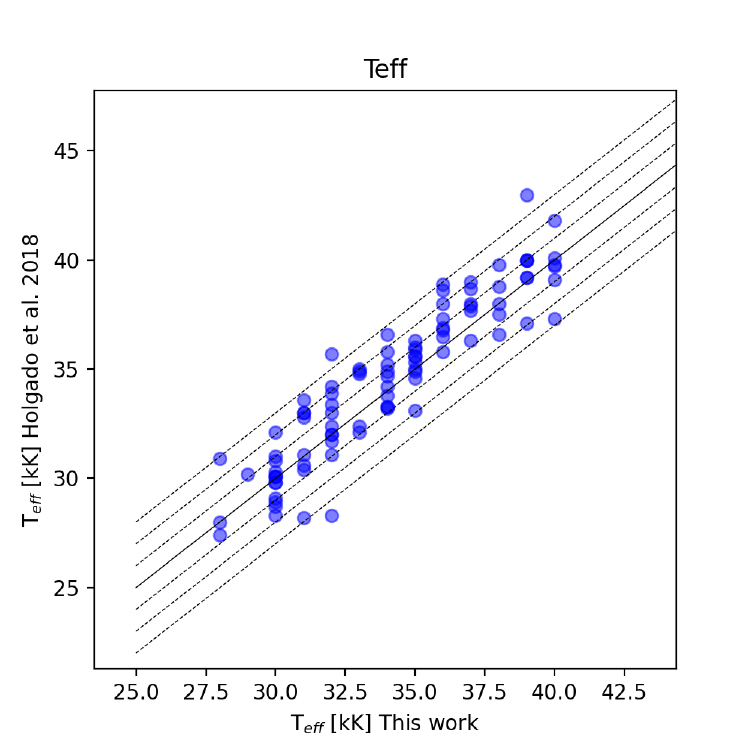}
\includegraphics[width=0.32\textwidth]{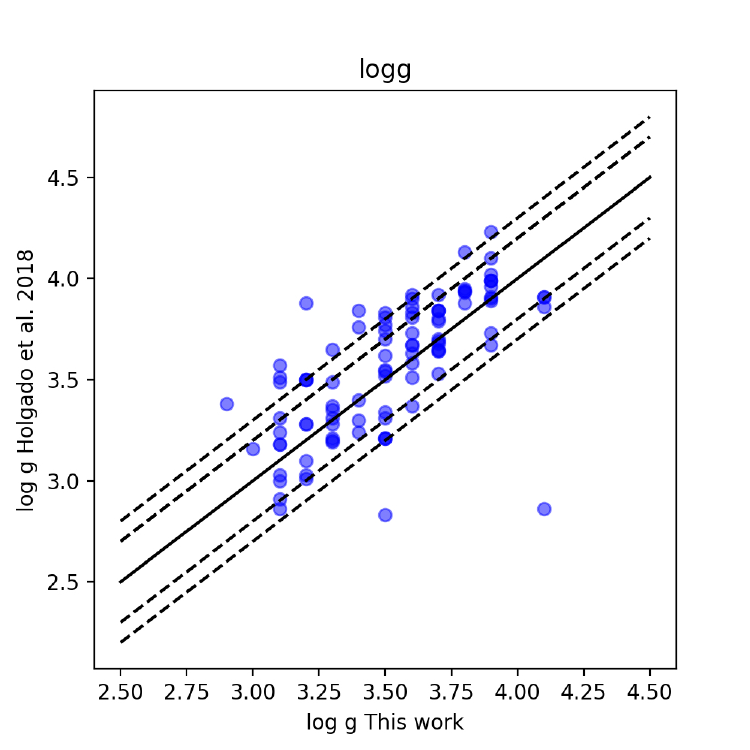}
\includegraphics[width=0.32\textwidth]{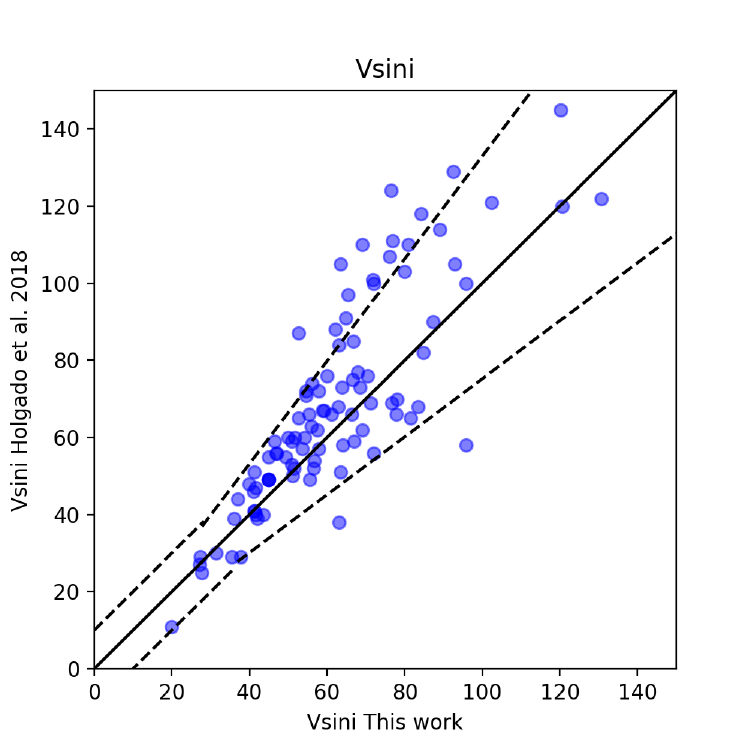}
\caption{\label{fig:comparison}Comparison of stellar parameters obtained by \textit{HiLineThere} in a fully automated analysis with those obtained by \cite{holgado2018} with a manual range and flux selection analysis for all the stars in their sample. The left figure illustrates the results in temperature; each of the dotted lines represents $1000$ [K]. The Center figure illustrates the result of the comparison for gravity; dotted lines  represent 0.2 and 0.3 $[\mathrm{dex}]$. Right figure, $V\sin\, i$ comparison, dotted line represent firstly $\pm 10 \mathrm{[km/s]} $ and later the $20\% $ of the value.}
\end{figure}

We have tested the automatic processes by uploading the sample studied in detail by \cite{holgado2018} and comparing their result with those obtained here; we present preliminary results in Figure \ref{fig:comparison}.

We are finding differences within the expected errors, $T_{\mathrm{eff}} \sim 1500 [\mathrm{K}]$  , $\log \, g \sim 0.3[\mathrm{dex}]$ and $V\sin\,i \sim 20\%$. It is important to highlight that all these stellar parameters, which include the $V\sin\,i$, RV, Temperature and gravity, were obtained entirely in a fully automatic way by using \textit{HiLineThere}.

%
%
\small  
%
\section*{Acknowledgements}
The Generalitat Valenciana partially supports this research under grant PROMETEO/2019/041, and  the Spanish Government Ministerio de Ciencia e Innovación and Agencia Estatal de Investigación under grant PGC2018-093741-B-C21/22 (MICIU/AEI/FEDER, UE) and grant PID2021-122397NB-C22 (MCIN/AEI/10.13039/501100011033/FEDER, UE).

%
\bibliographystyle{rubkek} 
\bibliography{rubkek}
\end{document}